\documentclass[preprint,5p,times,number]{elsarticle}

\usepackage{hyperref}
\usepackage{amsmath}
\usepackage{amssymb}
\usepackage{epsfig}
\usepackage{float}
\usepackage{bm}
\usepackage{epsfig}
\usepackage{soul}
\usepackage{color}
\usepackage{tikz}

\newcommand{\ri}{\mathrm{i}}
\newcommand{\re}{\mathrm{e}}
\newcommand{\rd}{\mathrm{d}}

\journal{Physics Letters A}

\begin{document}

\begin{frontmatter}

\title{Functional integral derivation of the kinetic equation\\of two-dimensional point vortices}

\author[label1]{Jean-Baptiste Fouvry}
\author[label2]{Pierre-Henri Chavanis}
\author[label1,label3]{Christophe Pichon}

\address[label1]{Institut d'Astrophysique de Paris and UPMC, CNRS (UMR 7095), 98 bis Boulevard Arago, 75014, Paris, France}
\address[label2]{Laboratoire de Physique Th\'eorique (IRSAMC), CNRS and UPS, Univ. de Toulouse, F-31062 Toulouse, France}
\address[label3]{Korea Institute of Advanced Studies (KIAS) 85 Hoegiro, Dongdaemun-gu, Seoul, 02455, Republic of Korea}

\begin{abstract}

We present a brief derivation of the kinetic equation describing the secular
evolution of point vortices in two-dimensional hydrodynamics, by relying on a functional integral formalism.
We start from Liouville's equation which describes the exact dynamics of a two-dimensional system of point vortices.
At the order ${1/N}$, the evolution of the system is characterised by the first two equations of the BBGKY hierarchy involving the system's ${1-}$body distribution function and its ${2-}$body correlation function.
Thanks to the introduction of auxiliary fields, these two evolution constraints may be rewritten as a functional integral. When functionally integrated over the ${2-}$body correlation function, this rewriting leads to a new constraint coupling the ${1-}$body distribution function and the two auxiliary fields. Once inverted, this constraint provides, through a new route, the closed non-linear kinetic equation satisfied by the ${1-}$body distribution function.
Such a method sheds new lights on the origin of these kinetic equations complementing the traditional derivation methods.
\end{abstract}


\end{frontmatter}

\section{Introduction}
\label{sec:introduction}

There exist beautiful analogies between stellar systems and two-dimensional (${2D}$)
vortices~\cite{houches}. Stellar systems and ${2D}$ point vortices undergo
two
successive types of relaxation. They first reach a quasistationary state (QSS)
due to a process of violent collisionless relaxation. The concept of violent
relaxation was introduced by Lynden-Bell~\cite{lb} in the case of stellar
systems described by the Vlasov equation and by Miller~\cite{miller} and Robert
and Sommeria~\cite{rs} in the case of ${2D}$ vortices described by the ${2D}$ Euler
equation (see~\cite{csr} for a description of the close link between these
two theories).
These QSSs correspond to galaxies in astrophysics~\cite{bt} or to large scale
vortices (like Jupiter's great red spot) in geophysical and astrophysical flows~\cite{bv}. On a longer (secular) timescale, ``collisions''\footnote{These
``collisions'' do not correspond to physical collisions but rather to - possibly
distant - encounters between the particles. They account for fluctuations due to
 finite${-N}$ effects, i.e., for the granularity of the system.}
between stars or between point vortices come into play  and drive the
system towards a statistical equilibrium state described by the Boltzmann
distribution. This statistical equilibrium state was conjectured  by Ogorodnikov~\cite{ogorodnikov} in the case of stellar
systems and by Onsager~\cite{onsager,esr} and Montgomery and Joyce~\cite{mj} in
the case of ${2D}$ point vortices. Actually, for collisional stellar systems such as
globular clusters the relaxation towards the Boltzmann statistical equilibrium
state is hampered by the evaporation of stars~\cite{spitzer} and by the
gravothermal catastrophe~\cite{antonov,lbw}. In the case of
${2D}$ point vortices, the statistical equilibrium state may present the
peculiarity to have a negative temperature as first noted by
Onsager~\cite{onsager}.

To understand the dynamical evolution of these systems, we need to develop a
kinetic theory. The collisionless evolution of stellar systems  is described by
the Vlasov~\cite{vlasov} equation that was first written by Jeans~\cite{jeans}
in astrophysics.\footnote{The kinetic theories of stellar
systems and neutral Coulombian plasmas have been developed in parallel (and often independently) by
astrophysicists and plasma physicists.} The
collisional evolution of stellar systems is usually
described by the Fokker-Planck equation introduced by Chandrasekhar~\cite{chandra1} or by the Landau~\cite{landau} equation. These
equations rely on a local approximation (as if the system were spatially
homogeneous) and neglect collective effects (i.e., the dressing of the stars by
their polarisation cloud). A gravitational Landau equation that takes
into account spatial inhomogeneity through the use of angle-action variables 
has been introduced in~\cite{landauaa,unified,chavanisaa}
 and a gravitational Balescu-Lenard equation that takes into account  spatial
inhomogeneity and collective effects has been introduced in~\cite{heyvaerts,quasistellar}. These equations have recently been
applied to stellar discs in~\cite{fouvry1,fouvry2,fouvry3}.

Exploiting the analogy between ${2D}$ vortices and stellar systems,  a
kinetic
theory of point vortices has been elaborated by Chavanis~\cite{kin}. The
collisionless evolution of point vortices is described by the ${2D}$ Euler equation.
When collective effects are neglected, the collisional evolution of  point
vortices is described by a Landau-type
equation~\cite{kin,cl,kinvortex}. A Balescu-Lenard-type equation taking
collective effects into account has been derived in~\cite{quasivortex,bbgkyvortex} for an axisymmetric distribution of point
vortices. It is equivalent to the one derived in~\cite{dn} in the
similar context of non-neutral plasmas.

One can understand the collisional evolution of stellar systems and ${2D}$ point
vortices heuristically by analogy with the Brownian motion. A star has a
diffusive motion due to the fluctuations of the gravitational force but it also
experiences a dynamical friction~\cite{chandrabrownian}.
Similarly, a point vortex has a diffusive motion due to the fluctuations of the
velocity field and also experiences a systematic drift~\cite{kin}. The
diffusion can be understood by considering the statistics of the gravitational
force created by a random distribution of stars~\cite{cn} or the statistics of
the velocity created by a random distribution of point vortices~\cite{cs}.
The dynamical friction experienced by a star and the
systematic drift experienced by a point vortex  can be
understood from a polarisation process and a linear response
theory (see~\cite{kandrupfriction} for
stellar systems and~\cite{drift} for point vortices). The friction
and drift coefficients are related to the diffusion coefficient by a form of
Einstein relation. Further analogies between the kinetic theory of stellar
systems, ${2D}$ vortices, and systems with long-range interactions in general are
discussed in~\cite{unified}.

There are many methods to derive kinetic equations for systems with long-range
interactions. The most popular are the BBGKY
hierarchy based on the Liouville equation (see~\cite{balescu,lenard} for
plasmas,~\cite{gilbert,severne,paper3bbgky,heyvaerts,chavanisaa} for stellar
systems and~\cite{kinvortex,bbgkyvortex}
for point vortices), the quasilinear theory based on the Klimontovich equation 
(see~\cite{klimontovich} for plasmas,~\cite{paper4quasi,quasistellar} for
stellar systems and~\cite{dn,quasivortex,kinvortex} for point vortices), and the
projection operator technique also based on the Liouville equation (see~\cite{kandrup} for stellar systems and~\cite{kin} for point vortices). One can
also derive kinetic equations from a field theory. This method was introduced
by Jolicoeur and Le Guillou~\cite{JolicoeurGuillou1989} to derive the homogeneous Balescu-Lenard
equation of plasma physics.
Recently, this method was generalised to stellar systems in~\cite{FouvryChavanisPichon2016} to derive the inhomogeneous Landau equation. Owing to
the analogy between stellar systems and ${2D}$ point vortices, it is of interest to
show how this method can be used to derive the Landau equation for axisymmetric
point vortices.

The present letter is organised as follows.
Section~\ref{sec:BBGKY} presents a brief derivation of the relevant BBGKY hierarchy in the context of the kinetic theory of ${2D}$ point vortices.
Section~\ref{sec:formalism} details the functional integral formalism introduced in~\cite{JolicoeurGuillou1989} and applied in~\cite{FouvryChavanisPichon2016} for inhomogeneous long-range systems.
Section~\ref{sec:application} illustrates how this formalism may be used to obtain the Landau equation describing the secular evolution of axisymmetric ${2D}$ point vortices.
Section~\ref{sec:discussion} discusses the limitations of our approach and its possible extensions.
Finally, section~\ref{sec:conclusion} wraps up.

\section{Derivation of the BBGKY hierarchy}
\label{sec:BBGKY}

In this section, we briefly recover the evolution equations describing the 
dynamics of point vortices and the associated BBGKY hierarchy. We consider a ${
2D }$ system made of $N$ point vortices of individual circulation ${ \gamma
\!=\! \Gamma_{\rm tot} / N }$. The individual dynamics of these vortices is
entirely described by the Kirchhoff-Hamilton equations which read~\cite{newton}:
\begin{equation}
\gamma \frac{\rd x_{i}}{\rd t} = \frac{\partial H}{\partial y_{i}} \;\;\; ; \;\;\;  \gamma \frac{\rd y_{i}}{\rd t} = - \frac{\partial H}{\partial x_{i}} \, ,
\label{Kirchoff_Hamilton_equations}
\end{equation}
where we introduced the coordinates ${ \bm{r} \!=\! (x , y) }$, as well as the Hamiltonian ${ H \!=\! \gamma^{2} \sum_{i < j} u_{ij} }$, where ${ u_{ij} \!=\! - 1/(2 \pi) \ln( |\bm{r}_{i} \!-\! \bm{r}_{j}| ) }$ is the potential of interaction between two vortices. We may now introduce the ${N-}$body probability distribution function (PDF) ${ P_{N} (\bm{r}_{1} , ... , \bm{r}_{N} , t) }$, which describes the probability of finding the vortex $1$ at position $\bm{r}_{1}$, vortex $2$ at position $\bm{r}_{2}$, etc. We normalise $P_{N}$ such that ${ \!\! \int \!\! \rd \bm{r}_{1} ... \rd \bm{r}_{N} \, P_{N} (\bm{r}_{1} , ... , \bm{r}_{N} , t) \!=\! 1 }$. The evolution of $P_{N}$ is then governed by Liouville's equation which reads
\begin{equation}
\frac{\partial P_{N}}{\partial t} \!+\! \gamma \sum_{i = 1}^{N} \bm{V}_{i} \!\cdot\! \frac{\partial P_{N}}{\partial \bm{r}_{i}} = 0 \, ,
\label{Liouville_equation}
\end{equation}
where we defined the velocity ${ \bm{V}_{i} \!=\! \sum_{j \neq i} \! \bm{V}_{ij} \!=\! \sum_{j \neq i} - \bm{e}_{z} \!\times\! \partial u_{ij} / \partial \bm{r}_{i} }$. Here, $\bm{V}_{ij}$ denotes the exact velocity induced by the vortex $j$ on the vortex $i$. We now introduce the reduced distribution functions (DF) $f_{n}$ as
\begin{equation}
f_{n} (\bm{r}_{1} , ... , \bm{r}_{n} , t) = \gamma^{n} \frac{N!}{(N \!-\! n)!} \!\! \int \!\! \rd \bm{r}_{n+1} ... \rd \bm{r}_{N} \, P_{N} (\bm{r}_{1} , ... , \bm{r}_{N} , t) \, .
\label{definition_fn}
\end{equation}
Integrating equation~\eqref{Liouville_equation} w.r.t. ${ (\bm{r}_{n+1} , ... , \bm{r}_{N}) }$, one obtains a BBGKY-like hierarchy of equations as
\begin{equation}
\frac{\partial f_{n}}{\partial t} \!+\! \sum_{ i = 1}^{n} \sum_{k = 1 , k \neq i}^{n} \gamma \bm{V}_{ik} \!\cdot\! \frac{\partial f_{n}}{\partial \bm{r}_{i}} \!+\! \sum_{i = 1}^{n} \!\! \int \!\! \rd \bm{r}_{n + 1} \, \bm{V}_{i , n+1} \!\cdot\! \frac{\partial f_{n+1}}{\partial \bm{r}_{i}} = 0 \, .
\label{BBGKY_fn}
\end{equation}
We are interested in the contributions arising from the correlations between particles, and therefore introduce the cluster representation of the DF. Indeed, we define the ${2-}$ and ${3-}$body correlation functions $g_{2}$ and $g_{3}$ as
\begin{align}
& \, f_{2} ( \bm{r}_{1} , \bm{r}_{2}) =  f_{1} (\bm{r}_{1}) f_{1} (\bm{r}_{2}) \!+\! g_{2} (\bm{r}_{1} , \bm{r}_{2}) \, ,   \nonumber
\\
& \, f_{3} (\bm{r}_{1} , \bm{r}_{2} , \bm{r}_{3})  = f_{1} (\bm{r}_{1}) f_{1} (\bm{r}_{2}) f_{1} (\bm{r}_{3})  \nonumber
\\
& \;\;\;\;\; \;\;\;\;\; \;\;\;\;\; \; \!+\! f_{1} (\bm{r}_{1}) g_{2} (\bm{r}_{2} , \bm{r}_{3}) \!+\! f_{1} (\bm{r}_{2}) g_{2} (\bm{r}_{1} , \bm{r}_{3}) \!+\! f_{1} (\bm{r}_{3}) g_{2} (\bm{r}_{1} , \bm{r}_{2})  \nonumber
\\
& \;\;\;\;\; \;\;\;\;\; \;\;\;\;\; \; \!+\! g_{3} (\bm{r}_{1} , \bm{r}_{2} , \bm{r}_{3} ) \, .
\label{definition_g2_g3}
\end{align}
It is then straightforward to check that one has the normalisations
\begin{align}
& \, \!\! \int \!\! \rd \bm{r}_{1} \,  f_{1} (\bm{r}_{1}) \!=\! \gamma N \;\; ; \;\;\;
\!\! \int \!\! \rd \bm{r}_{1} \rd \bm{r}_{2} \,  g_{2} (\bm{r}_{1} , \bm{r}_{2}) \!=\! - \gamma^{2} N \, ,   \nonumber
\\
& \, \!\! \int \!\! \rd \bm{r}_{1} \rd \bm{r}_{2} \rd \bm{r}_{3} \, g_{3} (\bm{r}_{1} , \bm{r}_{2} , \bm{r}_{3}) \!=\! 2 \gamma^{3} N  \, .
\label{normalisations_f_g}
\end{align}
Since the individual circulation scales like ${ \gamma \!\sim\! 1/N }$, one immediately has ${ |f_{1}| \!\sim\! 1 }$, ${ |g_{2}| \!\sim\! 1/ N }$, and ${ | g_{3} | \!\sim\! 1/N^{2} }$. In order to consider quantities of order ${1}$, we introduce the system's ${1-}$body DF $F$, and ${2-}$body correlation function $\mathcal{C}$ as
\begin{equation}
F = f_{1} \;\;\; ; \;\;\; \mathcal{C} = \frac{g_{2}}{\gamma} \, .
\label{definition_F_C}
\end{equation}
When truncated at the order ${ 1/N }$, one can easily show that the first two equations of the hierarchy from equation~\eqref{BBGKY_fn} become
\begin{equation}
\frac{\partial F}{\partial t} \!+\! \bigg[ \!\! \int \!\! \rd \bm{r}_{2} \, \bm{V}_{12} F (\bm{r}_{2}) \bigg] \!\cdot\! \frac{\partial F}{\partial \bm{r}_{1}} \!+\! \gamma \!\! \int \!\! \rd \bm{r}_{2} \, \bm{V}_{12} \!\cdot\! \frac{\partial \mathcal{C} (\bm{r}_{1} , \bm{r}_{2})}{\partial \bm{r}_{1}} = 0 \, , 
\label{BBGKY_1}
\end{equation}
and
\begin{align}
& \, \frac{1}{2} \frac{\partial \mathcal{C} (\bm{r}_{1} , \bm{r}_{2})}{\partial t} \!+\! \bigg[ \!\! \int \!\! \rd \bm{r}_{3} \, \bm{V}_{13} F (\bm{r}_{3}) \bigg] \!\cdot\! \frac{\partial C (\bm{r}_{1} , \bm{r}_{2})}{\partial \bm{r}_{1}} \!+\! \bm{V}_{12} \!\cdot\! \frac{\partial F}{\partial \bm{r}_{1}} F (\bm{r}_{2}) \nonumber
\\
& \,  \!+\! \bigg[ \!\! \int \!\! \rd \bm{r}_{3} \, \bm{V}_{13} \mathcal{C} (\bm{r}_{2} , \bm{r}_{3}) \bigg] \!\cdot\! \frac{\partial F}{\partial \bm{r}_{1}} \!+\! (1 \!\leftrightarrow\! 2) = 0 \, ,
\label{BBGKY_2}
\end{align}
where ${ (1 \!\leftrightarrow\! 2) }$ stands for the permutation of indices $1$ and $2$, and applies to all preceding terms.
The two equations~\eqref{BBGKY_1} and~\eqref{BBGKY_2} form a system of two coupled evolution equations involving $F$ and $\mathcal{C}$, and are at the centre of the upcoming functional integral formalism.

\section{Functional integral formalism}
\label{sec:formalism}

When introducing their application of the functional integral formalism in the context of classical kinetic theory,~\cite{JolicoeurGuillou1989} showed how two coupled evolution equations such as equations~\eqref{BBGKY_1} and~\eqref{BBGKY_2} may be rewritten under a functional form. As an illustration of this approach, let us consider a dynamical quantity $f$ depending on the time $t$ and defined on a phase space $\Gamma$. We assume that this quantity satisfies an evolution equation of the form ${ [ \partial_{t} \!+\! L ] f \!=\! 0 }$, where $L$ is a differential operator. Introducing an auxiliary field $\lambda$ defined on the same space as $f$, the evolution constraint on $f$ can be rewritten under the form (see~\cite{JolicoeurGuillou1989} and~\cite{FouvryChavanisPichon2016} for more details) 
\begin{equation}
1 = \!\! \int \!\! \mathcal{D} f \mathcal{D} \lambda \, \exp \bigg[ \ri \!\! \int \!\! \rd t \rd \Gamma \, \lambda \big[ \partial_{t} \!+\! L \big] f \bigg] \, .
\label{functional_definition}
\end{equation}
In equation~\eqref{functional_definition}, we define the action ${ S [F , \lambda] \!=\! \ri \!\! \int \!\! \rd t \rd \Gamma \, \lambda [\partial_{t} \!+\! L] f }$ as the argument of the exponential. One should finally note that the evolution equation satisfied by $f$ corresponds to the quantity by which the auxiliary field $\lambda$ is multiplied in the action.

When considering the two coupled equations~\eqref{BBGKY_1} and~\eqref{BBGKY_2} involving $F$ and $\mathcal{C}$, one may proceed to a similar transformation. Indeed, introducing two auxiliary fields ${ \lambda_{1} (t , \bm{r}_{1}) }$ and ${ \lambda_{2} (t , \bm{r}_{1} , \bm{r}_{2}) }$ respectively associated with $F$ and $\mathcal{C}$, equations~\eqref{BBGKY_1} and~\eqref{BBGKY_2} can be rewritten under the functional form
\begin{align}
1 = \!\! \int \!\! \mathcal{D} F \mathcal{D} \mathcal{C} \mathcal{D} \lambda_{1} \mathcal{D} & \lambda_{2} \exp \bigg\{  \ri \!\! \int \!\! \rd t \rd \bm{r}_{1} \, \lambda_{1} (A_{1} F \!+\! B_{1} \mathcal{C})   \nonumber
\\
& \, \!+\! \frac{\ri}{2} \!\! \int \!\! \rd t \rd \bm{r}_{1} \rd \bm{r}_{2} \, \lambda_{2} (A_{2} \mathcal{C} \!+\! D_{2} \mathcal{C} \!+\! S_{2}) \bigg\} \, .
\label{functional_initial}
\end{align}
In equation~\eqref{functional_initial}, we introduced the operators $A_{1}$, $B_{1}$, $A_{2}$, $D_{2}$, and $S_{2}$ as
\begin{align}
A_{1} F & \, = \bigg[ \frac{\partial }{\partial t} \!+\! \bigg[ \!\! \int \!\! \rd \bm{r}_{2} \, \bm{V}_{12} F (\bm{r}_{2}) \bigg] \!\cdot\! \frac{\partial }{\partial \bm{r}_{1}}  \bigg] \, F (\bm{r}_{1}) \, ,  \nonumber
\\
B_{1} \mathcal{C} & \, = \gamma \!\! \int \!\! \rd \bm{r}_{2} \, \bm{V}_{12} \!\cdot\! \frac{\partial \mathcal{C} (\bm{r}_{1} , \bm{r}_{2})}{\partial \bm{r}_{1}} \, , \nonumber
\\
A_{2} \mathcal{C} & \, = \bigg[ \frac{\partial }{\partial t} \!+\! \!\! \int \!\! \rd \bm{r}_{3} \, F (\bm{r}_{3}) \bigg[ \bm{V}_{13} \!\cdot\! \frac{\partial }{\partial \bm{r}_{1}} \!+\! \bm{V}_{23} \!\cdot\! \frac{\partial }{\partial \bm{r}_{2}} \bigg] \bigg] \, \mathcal{C} (\bm{r}_{1} , \bm{r}_{2}) \, , \nonumber
\\
D_{2} \mathcal{C} & \, = \bigg[ \!\! \int \!\! \rd \bm{r}_{3} \, \bm{V}_{13} \mathcal{C} (\bm{r}_{2} , \bm{r}_{3}) \bigg] \!\cdot\! \frac{\partial F}{\partial \bm{r}_{1}} \!+\! (1 \!\leftrightarrow\! 2) \, , \nonumber
\\
S_{2} & \, = F (\bm{r}_{2}) \bm{V}_{12} \!\cdot\! \frac{\partial F}{\partial \bm{r}_{1}} \!+\! (1 \!\leftrightarrow\! 2) \, .
\label{definition_operators}
\end{align}
The prefactor ${ 1/2 }$ in equation~\eqref{functional_initial} was only added for later convenience and does not play any role on the final expression of the evolution equations, since it was added as a global prefactor.
One may now rewrite the functional integral from equation~\eqref{functional_initial} under the form
\begin{align}
1 = \!\! \int \!\! \mathcal{D} F \mathcal{D} \mathcal{C} \mathcal{D} \lambda_{1} \mathcal{D} & \lambda_{2} \exp \bigg\{ \ri \!\! \int \!\! \rd t \rd \bm{r}_{1} \, \lambda_{1} (\bm{r}_{1}) \, A_{1} F (\bm{r}_{1})   \nonumber
\\
& \, \!+\! \frac{\ri}{2} \!\! \int \!\! \rd t \rd \bm{r}_{1} \rd \bm{r}_{2} \, \lambda_{2} (\bm{r}_{1} , \bm{r}_{2}) \, G (\bm{r}_{1} , \bm{r}_{2})    \nonumber
\\
& \, \!-\! \frac{\ri}{2} \!\! \int \!\! \rd t \rd \bm{r}_{1} \rd \bm{r}_{2} \, \mathcal{C} (\bm{r}_{1} , \bm{r}_{2}) \, E (\bm{r}_{1} , \bm{r}_{2}) \bigg\} \, ,
\label{functional_rewrite}
\end{align}
where it is important to note that all the dependences w.r.t. $\mathcal{C}$ were gathered in the prefactor of the third line.
In equation~\eqref{functional_rewrite}, we introduced the quantity ${ G (\bm{r}_{1} , \bm{r}_{2}) }$ as
\begin{equation}
G (\bm{r}_{1} , \bm{r}_{2}) = \bm{V}_{12} \!\cdot\! \bigg[ F (\bm{r}_{2}) \frac{\partial F}{\partial \bm{r}_{1}} \!-\! F (\bm{r}_{1}) \frac{\partial F}{\partial \bm{r}_{2}} \bigg] \, ,
\label{definition_G}
\end{equation}
where we used the relation ${ \bm{V}_{ij} \!=\! - \bm{V}_{ji} }$. In equation~\eqref{functional_rewrite}, we also introduced the quantity ${ E (\bm{r}_{1} , \bm{r}_{2}) }$ given by
\begin{align}
E (\bm{r}_{1} , \bm{r}_{2}) \!=\! & \, A_{2} \lambda_{2} (\bm{r}_{1} , \bm{r}_{2}) \!+\! \!\! \int \!\! \rd \bm{r}_{3} \, \bigg[\! \bm{V}_{13} \lambda_{2} (\bm{r}_{2} , \bm{r}_{3}) \!+\! \bm{V}_{23} \lambda_{2} (\bm{r}_{1} , \bm{r}_{3}) \!\bigg] \!\cdot\! \frac{\partial F}{\partial \bm{r}_{3}}   \nonumber
\\
& \,  \!+\! \gamma \bm{V}_{12} \!\cdot\! \bigg[ \frac{\partial \lambda_{1}}{\partial \bm{r}_{1}} \!-\! \frac{\partial \lambda_{1}}{\partial \bm{r}_{2}} \bigg] \, ,
\label{definition_E_full}
\end{align}
obtained thanks to integrations by parts. In order to invert the time derivative ${ \partial \mathcal{C}/\partial t }$ present in the term ${ \lambda_{2} A_{2} \mathcal{C} }$ from equation~\eqref{functional_initial}, we assumed ${ t \!\in\! [ 0 ; T ] }$, where $T$ is an arbitrary temporal bound, along with the boundary conditions ${ \mathcal{C} (t \!=\! 0) \!=\! 0 }$ (the system is initially uncorrelated), and ${ \lambda_{2} (T) \!=\! 0 }$ (we are free to impose a condition on $\lambda_{2}$). As presented in~\cite{FouvryChavanisPichon2016}, we will now neglect collective effects, i.e., neglect contributions associated with the term ${ D_{2} \mathcal{C} }$ in equation~\eqref{functional_initial}. As a consequence, equation~\eqref{definition_E_full} becomes
\begin{equation}
E (\bm{r}_{1} , \bm{r}_{2}) = A_{2} \lambda_{2} (\bm{r}_{1} , \bm{r}_{2}) \!+\! \gamma \bm{V}_{12} \!\cdot\! \bigg[ \frac{\partial \lambda_{1}}{\partial \bm{r}_{1}} \!-\! \frac{\partial \lambda_{1}}{\partial \bm{r}_{2}} \bigg] \, .
\label{definition_E}
\end{equation}
In order to obtain a closed kinetic equation involving $F$ only, the traditional approach would be to start from equation~\eqref{functional_initial} and proceed as follows. By functionally integrating equation~\eqref{functional_initial} w.r.t. $\lambda_{2}$, one gets a constraint of the form ${ (A_{2} \mathcal{C} \!+\! D_{2} \mathcal{C} \!+\! S_{2}) \!=\! 0 }$, which effectively couples $\mathcal{C}$ and $F$. This must then be inverted to give ${ \mathcal{C} \!=\! \mathcal{C} [F] }$. Using this substitution in equation~\eqref{functional_initial} and functionally integrating it w.r.t. $\lambda_{1}$, one finally obtains a kinetic equation involving $F$ only. This is the Landau equation (or Balescu-Lenard equation when collective effects are accounted for).
However, based on the rewriting from equation~\eqref{functional_rewrite},~\cite{JolicoeurGuillou1989} suggested a different strategy. By functionally integrating equation~\eqref{functional_rewrite} w.r.t. ${ \mathcal{C} }$, one gets a constraint of the form ${ E[F , \lambda_{1} , \lambda_{2}] \!=\! 0 }$. When inverted, this constraint leads to a relation of the form ${ \lambda_{2} \!=\! \lambda_{2} [F , \lambda_{1}] }$. Substituting this expression in equation~\eqref{functional_rewrite}, one then obtains a functional equation which only involves $F$ and $\lambda_{1}$. When functionally integrating this equation w.r.t. $\lambda_{1}$, one finally obtains a closed kinetic equation involving F only.
In~\cite{FouvryChavanisPichon2016}, we illustrated how this approach may be used in the context of inhomogeneous systems and recovered the inhomogeneous Landau equation.
In the present letter, we show how this same approach naturally applies in the context of axisymmetric systems of point vortices, when collectives effects are neglected, and recover the results from~\cite{kin,cl,kinvortex}.

\section{Application to  systems of vortices}
\label{sec:application}

\subsection{Axisymmetric systems}
\label{sec:axisymmetric}

\begin{figure}
\begin{center}
\begin{tikzpicture}[scale = 1.]
\newcommand{\subss}{\tiny} ; 
\pgfmathsetmacro{\circrada}{0.3} ; 
\pgfmathsetmacro{\circradb}{1.1} ;
\pgfmathsetmacro{\circradc}{1.9} ; 
\pgfmathsetmacro{\circradd}{2.7} ;
\pgfmathsetmacro{\circrade}{3.5} ;
\newcommand{\drawcircle}[1]{\draw [gray!70 ,dotted, line cap=round , dash pattern=on 0pt off 4 \pgflinewidth , thick] (0,0) circle [radius = #1] ;} ;
\drawcircle{\circrada} ; 
\drawcircle{\circradb} ; 
\drawcircle{\circradc} ; 
\drawcircle{\circradd} ;
\drawcircle{\circrade} ;
\pgfmathsetmacro{\xmin}{-3.7} ; \pgfmathsetmacro{\xmax}{3.7} ;
\pgfmathsetmacro{\ymin}{-3.7} ; \pgfmathsetmacro{\ymax}{3.7} ;
\draw [-> , thick] (\xmin , 0) -- (\xmax , 0) ; \draw (\xmax , 0) node[font = \normalsize , right] {$x$} ; 
\draw [-> , thick] (0 , \ymin) -- (0 , \ymax) ; \draw (0 , \ymax) node[font = \normalsize , above] {$y$} ; 
\pgfmathsetmacro{\xa}{1} ; \pgfmathsetmacro{\ya}{1} ;
\pgfmathsetmacro{\xb}{-1} ; \pgfmathsetmacro{\yb}{1} ;
\pgfmathsetmacro{\xc}{2.2} ; \pgfmathsetmacro{\yc}{0.8} ;
\pgfmathsetmacro{\xd}{-0.4} ; \pgfmathsetmacro{\yd}{3.1} ;
\pgfmathsetmacro{\xe}{-1.2} ; \pgfmathsetmacro{\ye}{2.1} ;
\pgfmathsetmacro{\xf}{-0.3} ; \pgfmathsetmacro{\yf}{-1.4} ;
\pgfmathsetmacro{\xg}{-2.3} ; \pgfmathsetmacro{\yg}{-0.4} ;
\pgfmathsetmacro{\xh}{-2.5} ; \pgfmathsetmacro{\yh}{-1.9} ;
\pgfmathsetmacro{\xi}{1.4} ; \pgfmathsetmacro{\yi}{-1.7} ;
\pgfmathsetmacro{\xj}{3.2} ; \pgfmathsetmacro{\yj}{-0.4} ;
\pgfmathsetmacro{\pointss}{0.04} ;
\pgfmathsetmacro{\arcss}{0.18} ; 
\newcommand{\getrad}[2]{
{sqrt(#1 * #1 + #2 * #2)}} ;
\newcommand{\drawvortex}[3]{
\draw [fill] (#1 , #2) circle [radius = \pointss] ; 
\draw (#1 , #2) node[font = \normalsize , below right] {#3} ;
\draw[->,semithick] (#1 - \arcss , #2) arc[radius= \arcss, start angle=180, end angle=450] ; 
};
\drawvortex{\xa}{\ya}{$\bm{r}_{\text{\subss 1}}$} ; 
\drawvortex{\xb}{\yb}{$\bm{r}_{\text{\subss 2}}$} ;
\drawvortex{\xc}{\yc}{} ;
\drawvortex{\xd}{\yd}{} ;
\drawvortex{\xe}{\ye}{} ;
\drawvortex{\xf}{\yf}{} ; 
\drawvortex{\xg}{\yg}{} ;
\drawvortex{\xh}{\yh}{} ;
\drawvortex{\xi}{\yi}{} ;
\drawvortex{\xj}{\yj}{} ;

\draw [semithick] (0 , 0) -- (\xa , \ya) ; 
\draw (\xa / 2 , \ya / 2) node[font = \normalsize , above] {$r_{\text{\subss 1}}$} ; 
\draw [->,semithick] ({sqrt(\xa * \xa + \ya * \ya)/2} , 0) arc[ radius = {sqrt(\xa * \xa + \ya * \ya)/2} , start angle = 0 , end angle = {atan2(\xa,\ya)}] ; \draw ( { (sqrt(\xa * \xa + \ya * \ya)/2) * cos(atan2(\xa,\ya)/2) } , { (sqrt(\xa * \xa + \ya * \ya)/2) * sin(atan2(\xa,\ya)/2) } )  node[font = \normalsize , right] {$\theta_{\text{\subss 1}}$} ; 
\pgfmathsetmacro{\offazi}{5} ;
\newcommand{\drawarrow}[5]{
\draw [-> , semithick] ({\getrad{#1}{#2}*cos(atan2(#1,#2)+#4)} , {\getrad{#1}{#2}*sin(atan2(#1,#2)+#4)}) arc[ radius = {\getrad{#1}{#2}} , start angle = {atan2(#1,#2)+#4} , end angle = {atan2(#1,#2)+#4+#3}] ;
\draw ({\getrad{#1}{#2}*cos(atan2(#1,#2)+#4+#3/2) + 0.1} , {\getrad{#1}{#2}*sin(atan2(#1,#2)+#4+#3/2) - 0.1}) node[font = \normalsize , above] {#5} ; 
};
\pgfmathsetmacro{\rada}{\getrad{\xa}{\ya}};
\pgfmathsetmacro{\radb}{\getrad{\xb}{\yb}};
\pgfmathsetmacro{\radc}{\getrad{\xc}{\yc}};
\pgfmathsetmacro{\radd}{\getrad{\xd}{\yd}};
\pgfmathsetmacro{\rade}{\getrad{\xe}{\ye}};
\pgfmathsetmacro{\radf}{\getrad{\xf}{\yf}};
\pgfmathsetmacro{\radg}{\getrad{\xg}{\yg}};
\pgfmathsetmacro{\radh}{\getrad{\xh}{\yh}};
\pgfmathsetmacro{\radi}{\getrad{\xi}{\yi}};
\pgfmathsetmacro{\radj}{\getrad{\xj}{\yj}};
\drawarrow{\xa}{\ya}{100 / \rada^(1.5)}{10 / \rada}{$\Omega_{\text{\subss 1}}$} ;
\drawarrow{\xb}{\yb}{100 / \radb^(1.5)}{10 / \radb}{} ; 
\drawarrow{\xc}{\yc}{100 / \radc^(1.5)}{10 / \radc}{} ;
\drawarrow{\xd}{\yd}{100 / \radd^(1.5)}{10 / \radd}{} ; 
\drawarrow{\xe}{\ye}{100 / \rade^(1.5)}{10 / \rade}{} ;
\drawarrow{\xf}{\yf}{100 / \radf^(1.5)}{10 / \radf}{} ; 
\drawarrow{\xg}{\yg}{100 / \radg^(1.5)}{10 / \radg}{} ;
\drawarrow{\xh}{\yh}{100 / \radh^(1.5)}{10 / \radh}{} ; 
\drawarrow{\xi}{\yi}{100 / \radi^(1.5)}{10 / \radi}{} ;
\drawarrow{\xj}{\yj}{100 / \radj^(1.5)}{10 / \radj}{} ; 
\end{tikzpicture}
\caption{\small{Illustration of the axisymmetric system of ${2D}$ point vortices considered in section~\ref{sec:application}. For clarity, only a subset of the vortices is represented: the considered system would have a much larger value of $N$. Each vortex has the same circulation $\gamma$. The mean flow is azimuthal and characterised by the angular velocity $\Omega$.
}}
\label{fig_vortex}
\end{center}
\end{figure}
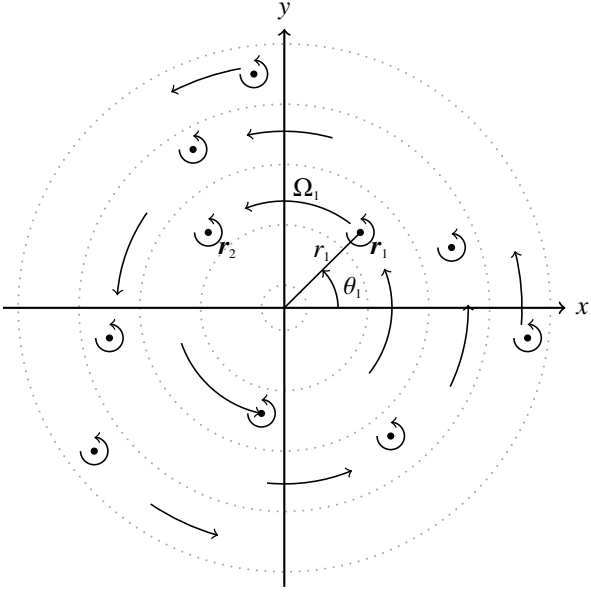

In order to obtain an explicit kinetic equation, we will now place ourselves in the simplified geometry of an axisymmetric distribution of point vortices, as illustrated in figure~\ref{fig_vortex}.
Such a geometry is straightforwardly a steady state of the ${2D}$ Euler equation, i.e., a stationary state of the collisionless dynamics. Introducing the polar coordinates ${ \bm{r} \!=\! (r , \theta) }$, one can assume that $F$, $\mathcal{C}$, and their associated auxiliary fields have the dependences
\begin{align}
F (\bm{r}_{1}) = F (r_{1}) \;\;\; & ; \;\;\; \mathcal{C} (\bm{r}_{1} , \bm{r}_{2}) = \mathcal{C} (r_{1} , r_{2} , \theta_{1} \!-\! \theta_{2}) \, ,  \nonumber
\\
\lambda_{1} (\bm{r}_{1}) = \lambda_{1} (r_{1}) \;\;\; & ;  \;\;\; \lambda_{2} (\bm{r}_{1} , \bm{r}_{2}) = \lambda_{2} (r_{1} , r_{2} , \theta_{1} \!-\! \theta_{2}) \, .
\label{dependence_axi}
\end{align}
Within the same simplified geometry, the mean axisymmetric flow satisfies
\begin{equation}
\!\! \int \!\! \rd \bm{r}_{2} \, \bm{V}_{12} F (\bm{r}_{2}) = \Omega (r_{1}) \, r_{1} \, \bm{e}_{\theta} \; ; \; \Omega (r_{1}) = \frac{1}{r_{1}^{2}} \!\! \int_{0}^{r_{1}} \!\!\!\! \rd r_{2} \, r_{2} \, F (r_{2}) \, ,
\label{definition_Omega}
\end{equation}
where ${ \Omega (r_{1}) }$ is the local angular velocity.
Thanks to these assumptions, one can rewrite the operators introduced in equation~\eqref{functional_rewrite} under a simpler form. Indeed, one has
\begin{equation}
A_{1} F = \frac{\partial F}{\partial t} \, .
\label{simpler_A1}
\end{equation}
Similarly, the term ${ G (\bm{r}_{1} , \bm{r}_{2}) }$ from equation~\eqref{definition_G} now reads
\begin{equation}
G (\bm{r}_{1} , \bm{r}_{2}) = \frac{1}{r_{1} r_{2}} \bigg[ r_{2} \frac{\partial u_{12}}{\partial \theta_{1}} F (r_{2}) \frac{\partial F}{\partial r_{1}} \!+\! r_{1} \frac{\partial u_{21}}{\partial \theta_{2}} F (r_{1}) \frac{\partial F}{\partial r_{2}} \bigg] \, .
\label{simpler_G}
\end{equation}
Finally, the constraint ${ E (\bm{r}_{1} , \bm{r}_{2}) }$ from equation~\eqref{definition_E} becomes
\begin{align}
E (\bm{r}_{1} , \bm{r}_{2}) = & \, \frac{\partial \lambda_{2}}{\partial t} \!+\! \Omega (r_{1}) \frac{\partial \lambda_{2}}{\partial \theta_{1}} \!+\! \Omega (r_{2}) \frac{\partial \lambda_{2}}{\partial \theta_{2}}  \nonumber
\\
& \, \!+\! \gamma \bigg[ \frac{1}{r_{1}} \frac{\partial u_{12}}{\partial \theta_{1}} \frac{\partial \lambda_{1}}{\partial r_{1}} \!+\! \frac{1}{r_{2}} \frac{\partial u_{21}}{\partial \theta_{2}} \frac{\partial \lambda_{1}}{\partial r_{2}} \bigg]  \, .
\label{simpler_E}
\end{align}

\subsection{Inverting the constraint}
\label{sec:constraint}

In order to invert equation~\eqref{simpler_E}, we rely on Bogoliubov ansatz (adiabatic approximation) by assuming that the fluctuations (such as $\mathcal{C}$ and $\lambda_{2}$) evolve on a much shorter timescale than the mean dynamical quantities (such as $F$ and $\lambda_{1}$). As a consequence, on the timescale for which $\lambda_{2}$ evolves, one can assume $F$ and $\lambda_{1}$ to be frozen, while on the timescale of secular evolution, one can assume $\lambda_{2}$ to be equal to the asymptotic value associated with the current values of $F$ and $\lambda_{1}$.

We introduce the azimuthal Fourier transform as
\begin{align}
f (r_{1} , r_{2} , \theta_{1} \!-\! \theta_{2}) & \, = \sum_{n} \re^{\ri n (\theta_{1} - \theta_{2})} f_{n} (r_{1} , r_{2})  \, ,\nonumber
\\
 f_{n} (r_{1} , r_{2}) & \, = \frac{1}{2 \pi} \!\! \int \!\! \rd \theta \, \re^{- \ri n \theta} \, f (r_{1} , r_{2} , \theta) \, .
\label{definition_TF_angles}
\end{align}
Multiplying equation~\eqref{simpler_E} by ${ 1/(2 \pi)^{2} \re^{\ri n (\theta_{1} - \theta_{2})} }$ and integrating w.r.t. $\theta_{1}$ and $\theta_{2}$, we obtain
\begin{equation}
\frac{\partial \lambda_{-n}}{\partial t} - \ri n \Delta \Omega \lambda_{-n}   - \gamma \ri n u_{n}^{*} (r_{1} , r_{2}) \bigg[ \frac{1}{r_{1}} \frac{\partial \lambda_{1}}{\partial r_{1}} \!-\! \frac{1}{r_{2}} \frac{\partial \lambda_{1}}{\partial r_{2}} \bigg]  = 0  \, ,
\label{E_TF}
\end{equation}
where we used the notations ${ \lambda_{-n} \!=\! \lambda_{-n} (r_{1} , r_{2}) }$, ${ \Delta \Omega \!=\! \Omega (r_{1}) \!-\! \Omega (r_{2}) }$, and relied on the fact that ${ u_{n} (r_{2} , r_{1}) \!=\! u_{-n} (r_{1} , r_{2}) \!=\! u_{n}^{*} (r_{1} , r_{2}) }$. Thanks to the boundary condition ${ \lambda_{2} (T) \!=\! 0 }$ used in equation~\eqref{definition_E_full}, equation~\eqref{E_TF} can straightforwardly be solved as
\begin{equation}
\lambda_{- n} (t) = - \gamma u_{n}^{*} (r_{1} , r_{2}) \bigg[ \frac{1}{r_{1}} \frac{\partial \lambda_{1}}{\partial r_{1}} \!-\! \frac{1}{r_{2}} \frac{\partial \lambda_{1}}{\partial r_{2}} \bigg] \frac{1 \!-\! \re^{\ri n \Delta \Omega  (t - T)}}{\Delta \Omega} \, .
\label{solution_E}
\end{equation}
In order to consider only the forced regime of evolution, we now assume that the arbitrary temporal bound $T$ is large compared to the time $t$, so that we place ourselves in the limit ${ T \!\to\! + \infty }$. We recall the formula
\begin{equation}
\lim\limits_{T \to + \infty} \frac{\re^{\ri T \Delta \omega} \!-\! 1}{\Delta \omega} = \ri \pi \delta_{\rm D} (\Delta \omega) \, ,
\label{formula_asymp}
\end{equation}
so that equation~\eqref{solution_E} immediately gives
\begin{equation}
\lim\limits_{T \to + \infty} \!\! \lambda_{-n} (t) \!=\! - \ri \pi \gamma \frac{n}{|n|} u_{n}^{*} (r_{1} , r_{2})  \bigg[\! \frac{1}{r_{1}} \frac{\partial \lambda_{1}}{\partial r_{1}} \!-\! \frac{1}{r_{2}} \frac{\partial \lambda_{2}}{\partial r_{2}} \!\bigg] \, \delta_{\rm D} (\Omega (r_{1}) \!-\! \Omega (r_{2})) \, ,
\label{solution_E_limit}
\end{equation}
where we used the property ${ \delta_{\rm D} (\alpha x) \!=\! \delta_{\rm D} (x) / |\alpha| }$.
Equation~\eqref{solution_E_limit} illustrates how the Bogoliubov ansatz allowed us to invert the constraint ${ E [F , \lambda_{1} , \lambda_{2}] \!=\! 0 }$ from equation~\eqref{definition_E} so as to obtain ${ \lambda_{2} \!=\! \lambda_{2} [F , \lambda_{1}] }$.

\subsection{Recovering the Landau collision operator}
\label{sec:recoverLandau}

We now substitute the inverted expression from equation~\eqref{solution_E_limit} into equation~\eqref{functional_rewrite}, which then only depends on $F$ and $\lambda_{1}$. The remaining action term ${ S [F , \lambda_{1}] }$ takes the form
\begin{equation}
S \![\!F , \lambda_{1}\!] \!\!=\!\! \ri \!\!\! \int \!\!\! \rd t \rd \bm{r}_{1} \lambda_{1} \!(\!\bm{r}_{1}\!) A_{1} F (\!\bm{r}_{1}\!) + \frac{\ri}{2} \!\!\! \int \!\!\! \rd t \rd \bm{r}_{1} \rd \bm{r}_{2} \lambda_{2} [\! F \!,\! \lambda_{1} \!] G (\!\bm{r}_{1} \!,\! \bm{r}_{2}\!) \, .
\label{initial_S}
\end{equation}
Thanks to the expressions of $A_{1}$ and $G$ from equations~\eqref{simpler_A1} and~\eqref{simpler_G}, and using the Fourier transform introduced in equation~\eqref{definition_TF_angles}, equation~\eqref{initial_S} takes the form
\begin{align}
& \!  S [F , \lambda_{1}] = \ri \!\! \int \!\! \rd t  \rd \bm{r}_{1} \, \lambda_{1} (\bm{r}_{1}) \frac{\partial F}{\partial t} \!-\! \frac{\ri}{2} (2 \pi)^{2} \!\! \int \!\! \rd t \rd r_{1} \rd r_{2}   \nonumber
\\
& \!\! \times \sum_{n} n \text{Im} \bigg[\! \lambda_{-n} (r_{1} , r_{2}) \, u_{n} (r_{1} , r_{2}) \!\bigg] \bigg[\! r_{2} F (r_{2}) \frac{\partial F}{\partial r_{1}} \!-\! r_{1} F (r_{1}) \frac{\partial F}{\partial r_{2}} \!\bigg] \, .
\label{rewrite_S}
\end{align}
Thanks to equation~\eqref{solution_E_limit}, we immediately have
\begin{align}
\text{Im} \bigg[ \lambda_{-n} (r_{1} , r_{2}) u_{n} (r_{1} , r_{2}) \bigg] = & \, - \pi \gamma \frac{n}{|n|} | u_{n} (r_{1} , r_{2}) |^{2}  \delta_{\rm D} (\Omega (r_{1}) \!-\! \Omega (r_{2})) \nonumber
\\
& \, \times \bigg[ \frac{1}{r_{1}} \frac{\partial \lambda_{1}}{\partial r_{1}} \!-\! \frac{1}{r_{2}} \frac{\partial \lambda_{2}}{\partial r_{2}} \bigg] \, .
\label{final_Im}
\end{align}
Introducing the notation ${ \chi (r_{1} , r_{2}) \!=\! \sum_{n} |n| |u_{n} (r_{1} , r_{2}) |^{2} }$, equation~\eqref{rewrite_S} then becomes
\begin{align}
& \, S [F , \lambda_{1}] = \ri \!\! \int \!\! \rd t  \rd \bm{r}_{1} \, \lambda_{1} (\bm{r}_{1}) \frac{\partial F}{\partial t} \!+\! \frac{\ri}{2} (2 \pi)^{2} \pi \gamma \!\! \int \!\! \rd t \rd r_{1} \rd r_{2} \, \chi (r_{1} , r_{2})  \nonumber
\\
& \, \times \delta_{\rm D} (\Omega (r_{1}) \!-\! \Omega (r_{2}))  \bigg[\! \frac{1}{r_{1}} \frac{\partial \lambda_{1}}{\partial r_{1}} \!-\! \frac{1}{r_{2}} \frac{\partial \lambda_{1}}{\partial r_{2}} \!\bigg] \bigg[\! r_{2} F (r_{2}) \frac{\partial F}{\partial r_{1}} \!-\! r_{1} F (r_{1}) \frac{\partial F}{\partial r_{2}} \!\bigg] \, .
\label{rewrite_S_II}
\end{align}
The final step of the calculation is to rewrite the second term of equation~\eqref{rewrite_S_II} under the form ${ \!\! \int \!\! \rd t \rd \bm{r}_{1} \lambda_{1} (\bm{r}_{1}) ... }$. Using an integration by parts, and accordingly permuting the indices ${ 1 \!\leftrightarrow\! 2 }$, this is a straightforward calculation. Finally, when changing the integration domain, one has to rely on the property
\begin{equation}
\!\! \int \!\! \rd r_{1} \, f (r_{1}) =  \!\! \int \!\! \rd \bm{r}_{1} \, \frac{1}{2 \pi} \frac{1}{r_{1}} f (r_{1}) \, .
\label{change_integration_domain}
\end{equation}
After calculation, equation~\eqref{rewrite_S_II} can easily be rewritten as
\begin{align}
S [F , \lambda_{1}] = &  \ri \!\! \int \!\! \rd t \rd \bm{r}_{1} \, \lambda_{1} (\bm{r}_{1}) \bigg\{ \frac{\partial F}{\partial t}   \nonumber
\\
& \, \!-\! 2 \pi^{2} \gamma \frac{1}{r_{1}} \frac{\partial }{\partial r_{1}} \bigg[ \!\! \int \!\! \rd r_{2} \, r_{2} \, \chi (r_{1} , r_{2}) \,  \delta_{\rm D} (\Omega (r_{1}) \!-\! \Omega (r_{2}))   \nonumber
\\
& \, \times \bigg[ \frac{1}{r_{1}} \frac{\partial }{\partial r_{1}} \!-\! \frac{1}{r_{2}} \frac{\partial }{\partial r_{2}} \bigg] F (r_{1}) F (r_{2}) \bigg] \bigg\} \, . 
\label{rewrite_S_III}
\end{align}
By integrating functionally equation~\eqref{rewrite_S_III} w.r.t. $\lambda_{1}$, one finally obtains a closed form expression for the kinetic equation as
\begin{align}
\frac{\partial F}{\partial t} = \, 2 \pi^{2} \gamma \frac{1}{r_{1}} \frac{\partial }{\partial r_{1}} \bigg[ \!\! \int & \,  \!\! \rd r_{2} \, r_{2} \, \chi (r_{1} , r_{2}) \, \delta_{\rm D} (\Omega (r_{1}) \!-\! \Omega (r_{2}))   \nonumber
\\
& \, \times \bigg[ \frac{1}{r_{1}} \frac{\partial }{\partial r_{1}} \!-\! \frac{1}{r_{2}} \frac{\partial }{\partial r_{2}} \bigg] F (r_{1}) F (r_{2}) \bigg] \, . 
\label{final_Landau}
\end{align}
Using the functional integral formalism presented in
section~\ref{sec:formalism}, we were therefore able to recover the collisional
secular evolution of an axisymmetric system of point vortices, when collective
effects are neglected, in full agreement with what was obtained
in~\cite{kin,cl,kinvortex}. We refer to~\cite{cl} for a discussion of the properties of equation~\eqref{final_Landau}, its numerical resolution, and some physical applications.

\section{Discussion}
\label{sec:discussion}

In this section, we discuss the assumptions made in our paper to obtain the
kinetic equation~\eqref{final_Landau}, and how such assumptions can be
overpassed.

\subsection{Collective effects}

We neglected collective effects. In principle, collective
effects could be taken into account in our functional integral formalism at the
price of more complicated calculations. However, inverting equation~\eqref{definition_E_full}, when collective effects are accounted for, does not
appear as straightforward. These effects can be taken into account in the quasilinear and BBGKY formalisms~\cite{quasivortex,bbgkyvortex,dn}. This leads to the Balescu-Lenard equation
[see Eq. (63) in~\cite{bbgkyvortex}] instead of the Landau equation~\eqref{final_Landau}. It is found that
collective effects do not alter the physical structure of the kinetic equation.
One only has to replace the Fourier transform of the bare potential of
interaction ${ u_{n} (r_{1} , r_{2}) }$ by a dressed potential of interaction. This changes
the form of the function ${ \chi(r_{1} , r_{2}) }$ in equation~\eqref{final_Landau} without changing the structure of this equation.

In the case of plasmas, collective effects~\cite{balescu,lenard} are
responsible for Debye shielding (a charge is surrounded by a cloud of opposite
charges that shields the Coulombian interaction) and they regularise the
large-scale logarithmic divergence appearing in the Landau equation~\cite{landau} when they are ignored. In the case of stellar systems, collective
effects~\cite{heyvaerts,quasistellar} are responsible for anti-shielding (a star
is surrounded by a cloud of stars that increases the gravitational interaction) and they can reduce the
relaxation time by several orders of magnitude with respect to the case where
they are not taken into account, as shown in~\cite{fouvry2} for stellar discs.
In the case of ${2D}$
point vortices, the importance of collective effects is more difficult to
estimate. Collective effects are less crucial than in plasma physics since the
Landau equation of point vortices~\eqref{final_Landau} that ignores them is
well-behaved mathematically
(it does not present any divergence). On the other hand, when 
a test particle approach and a (thermal) bath approximation are implemented, one
can show that collective effects
become negligible in the expression of the diffusion and drift coefficients
entering in the Fokker-Planck equation~\cite{quasivortex,bbgkyvortex}. Finally,
we emphasise that what really matters in the kinetic equation~\eqref{final_Landau} is the condition of resonance encapsulated in the $\delta_{\rm D}$ function. This
condition implies that the kinetic equation reduces to ${ \partial F / \partial t \!=\! 0 }$ when the profile of angular velocity is monotonic, which is the generic case
for Euler stable axisymmetric flows.\footnote{When the profile of angular velocity is initially non-monotonic, but nevertheless dynamically (Euler) stable, one can show that the effect of distant collisions between point vortices is precisely to make it become monotonic~\citep{cl}.}
This is independent of whether or not
collective effects are taken into account. For an axisymmetic distribution
of point vortices with a monotonic profile of angular velocity, the relaxation
time towards the Boltzmann distribution is then longer than ${ N t_{\rm
dyn} }$~\cite{quasivortex,bbgkyvortex}, as it is due to higher order correlations.

\subsection{Axisymmetric flows}

We restricted ourselves to the case of axisymmetric flows in which the
point vortices follow circular orbits. In terms of mathematical simplicity, this
situation is the counterpart of spatially homogeneous self-gravitating
systems in which the stars have rectilinear trajectories. The difference between
these two systems comes from the fact that, in an axisymmetric flow, a vortex
has the tendency to rotate due to the influence of the other vortices while, in
a homogeneous medium, a star has the tendency to follow a straight line due to
its inertia. A formal kinetic equation of point vortices
valid for arbitrary flows has been obtained in~\cite{kin} [see Eq. (128)] and
confirmed in~\cite{sano,kinvortex}. Explicit approximate expressions of this
formal kinetic equation for non-axisymmetric flows have been obtained in~\cite{kin} [see Eq. (137)] and in~\cite{yatsu} under different assumptions. It would be interesting to
develop a rigorous kinetic theory of
point vortices for arbitrary flows by introducing the analogue of the 
angle-action
variables used to treat spatially inhomogeneous stellar systems~\cite{heyvaerts,quasistellar}. It may also be interesting to consider the
kinetic theory of point vortices on a sphere where it can have potential
applications to
geophysical flows (see, e.g.,~\cite{herbert} for the development of the
Miller-Robert-Sommeria~\cite{miller,rs} statistical theory on a sphere).

\subsection{Multi-species system}

We have considered a single species point vortex gas. The kinetic theory of
point vortices can be generalised
to the case where the vortices have different circulations~\cite{dubin,cl}.\footnote{The equilibrium statistical mechanics of a
multi-species gas of point vortices is treated extensively in~\cite{virialvortex}.}
When the circulations have the same sign, the validity of the kinetic equation is
the same as the one derived in this paper. However, when vortices have
different signs, the situation is more complicated. First of all, in generic situations,
the system is not axisymmetric but consists of two large vortices (macroscopic
dipole), one blob with a positive circulation and one blob with a negative
circulation, or in three large vortices (macroscopic tripole), one blob with a
positive circulation surrounded by two blobs of negative circulation (or the
opposite). In that case, the kinetic theory must be generalised to
non-axisymmetric flows. Point vortices can also form microscopic dipoles, pairs ${ (+,-) }$ of
positive and negative vortices, that have a ballistic motion and escape
to infinity. In that case, there is no equilibrium state. They
can also form microscopic tripoles ${ (+,-,+) }$ or ${ (-,+,-) }$. These structures,
corresponding to nontrivial correlations, are not taken into account in the
kinetic theory developed in our paper. The formation of these structures may be
negligible in the thermodynamic limit ${ N \!\rightarrow\! + \infty }$ considered here, but these structures may initially be present in the flow. Similarly, the presence of vortex pairs ${ (+,+) }$ or ${ (-,-) }$, similar to binary stars in astrophysics, is not taken into account here. Kinetic equations for a vortex gas viewed as a coupling, via the Liouville equation, between monopoles, dipoles and tripoles have been derived in~\cite{marmanis,newtonmezic}. Kinetic theory of
three-body collisions (dipoles hitting monopoles) with application to the
context of 2D decaying turbulence has also been developed in~\cite{sirechavanis}. There is finally also the possibility that point vortices split in three vortices (offsprings of a point vortex), a process reverse to the three point vortex collapse, offering the possibility of a statistical mechanics approach with a varying number of vortices~\cite{leoncini}.

\section{Conclusion}
\label{sec:conclusion}

Relying on the functional integral formalism introduced in~\cite{JolicoeurGuillou1989}, we illustrated how one may use this approach to derive the kinetic equation describing the long-term evolution of an axisymmetric distribution of point vortices when collective effects are neglected. We believe that such calculations allow for additional insights on the origin of these kinetic equations and complement the usual methods of derivation.
Rewriting kinetic theories for ${N-}$body systems with long-range interactions under a functional form allows for insightful connections with standard field theory methods such as the Martin-Siggia-Rose functional method for classical stochastic systems in the Jensen path integral formulation~\citep{MartinSiggiaRose1973}.
A next step of the current approach would be to show how the same methodology may be used when collective effects or higher order correlation terms are accounted for. This will be the subject of a future work.

\section*{Acknowledgements}
\label{sec:acknowledgements}

JBF, PHC and CP thank the CNRS Inphyniti program for funding.
JBF and CP thank the Korea Institute for Advanced Study  and the CFHT for hospitality.
This work is partially supported by the Spin(e) grants ANR-13-BS05-0005 of the French Agence Nationale de la Recherche
(\texttt{http://cosmicorigin.org}).

\section*{References}

\bibliographystyle{elsarticle-num-names}
\bibliography{references.bib}

\end{document}